\documentclass{eas}
\usepackage{graphicx}
\begin{document}

%%-----------------------------
%%      the top matter
%%-----------------------------
\title{Energetic processes revealed by spectrally resolved high-$J$ CO lines in low-mass star-forming regions with Herschel-HIFI} 
\runningtitle{High-$J$ CO lines in low-mass star-forming regions with Herschel}
\author{Umut~A.~Y{\i}ld{\i}z}\address{Leiden Observatory, Leiden University, PO Box 9513, 2300 RA Leiden, The Netherlands; \email{yildiz@strw.leidenuniv.nl}}
\author{Ewine~F.~van~Dishoeck}\sameaddress{1}\secondaddress{Max Planck Institut f\"{u}r Extraterrestrische Physik, 85748 Garching, Germany}
%\author{E.~F.~van~Dishoeck}\sameaddress{1}\secondaddress{Max Planck Institut f\"{u}r Extraterrestrische Physik, Giessenbachstrasse 1, 85748 Garching, Germany}
\author{Lars~E. Kristensen}\sameaddress{1}
\author{Ruud~Visser}\sameaddress{1}
\author{Greg~Herczeg}\sameaddress{2}
\author{Tim~A.~van~Kempen}\sameaddress{1}
%\author{T.~A.~van~Kempen}\address{Harvard-Smithsonian Center for Astrophysics, 60 Garden Street, MS 42, Cambridge, MA 02138, USA}
\author{Jes~K.~J{\o}rgensen}\address{Centre for Star and Planet Formation, Natural History Museum of Denmark, University of Copenhagen, {\O}ster Voldgade 5-7, DK-1350 Copenhagen K., Denmark}
\author{Michiel~R.~Hogerheijde}\sameaddress{1}
\author{the WISH Team}

\begin{abstract}
{\textit{Herschel}-HIFI observations of high-$J$ lines (up to $J_\mathrm{u}$=10) 
of $^{12}$CO, $^{13}$CO and C$^{18}$O are presented toward three deeply embedded 
low-mass protostars in NGC1333. 
%The spectrally-resolved HIFI data are complemented by ground-based observations of lower-$J$ CO and 
%isotopologue lines. 
The observations show several energetic components including shocked and quiescent gas. Radiative transfer models are used 
to quantify the C$^{18}$O envelope abundance which require a jump in the abundance at an evaporation temperature, $T_{\rm ev}$ $\sim$25~K, providing new direct evidence 
of a CO ice evaporation zone around protostars.
The abundance in the outermost part of the envelope, $X_{0}$, is within the canonical value of 2$\times$10$^{-4}$; however the inner abundance, $X_{\rm in}$, is found around a factor of 3-5 lower than $X_{0}$.
}
%The inner envelope abundance is, however, lower than the canonical value of 2$\times$10$^{-4}$, however, the outermost part of the envelope .}

\end{abstract}
\maketitle
%%-----------------------------
%%      your text
%%-----------------------------
\section{Introduction}
To understand the physical and chemical evolution of low-mass protostars, in particular the relative
importance of radiative heating and shocks in their energy budget,
spectrally resolved observations are required that can separate these components. 
The Heterodyne Instrument for the Far-Infrared (HIFI) on
\textit{Herschel} opens up the possibility to obtain high resolution 
data from high-frequency lines that are sensitive to gas
temperatures up to several hundred Kelvin. 
Three well-known low-mass protostars (IRAS~2A, 4A and 4B) are studied as part of the \textit{Water In Star-forming regions 
with Herschel} (WISH) key program (\cite{wish11}). 

%%-----------------------------
\section{Results}
\begin{figure}[h]
\begin{center}
\includegraphics[scale=0.18]{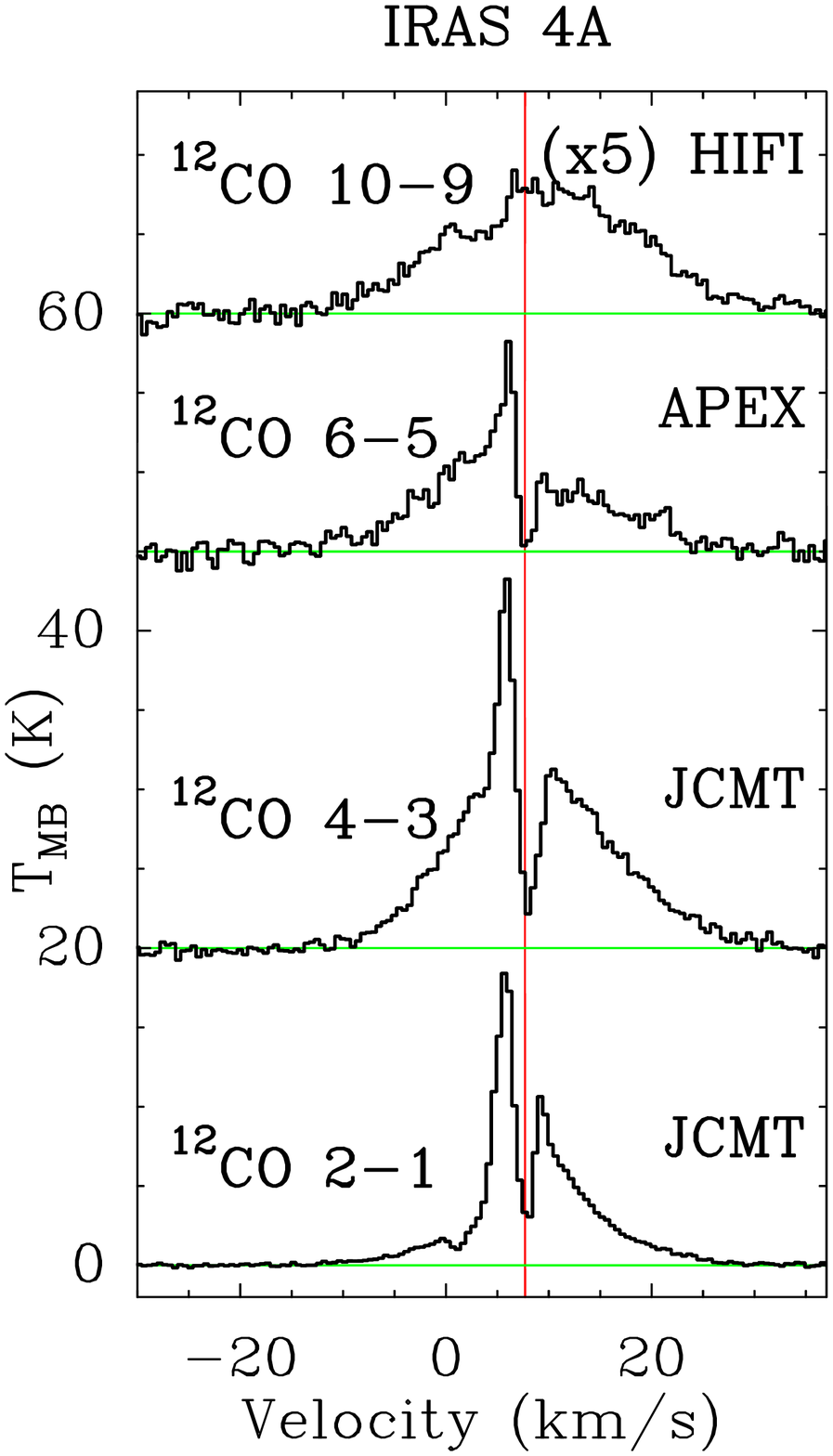}
\includegraphics[scale=0.18]{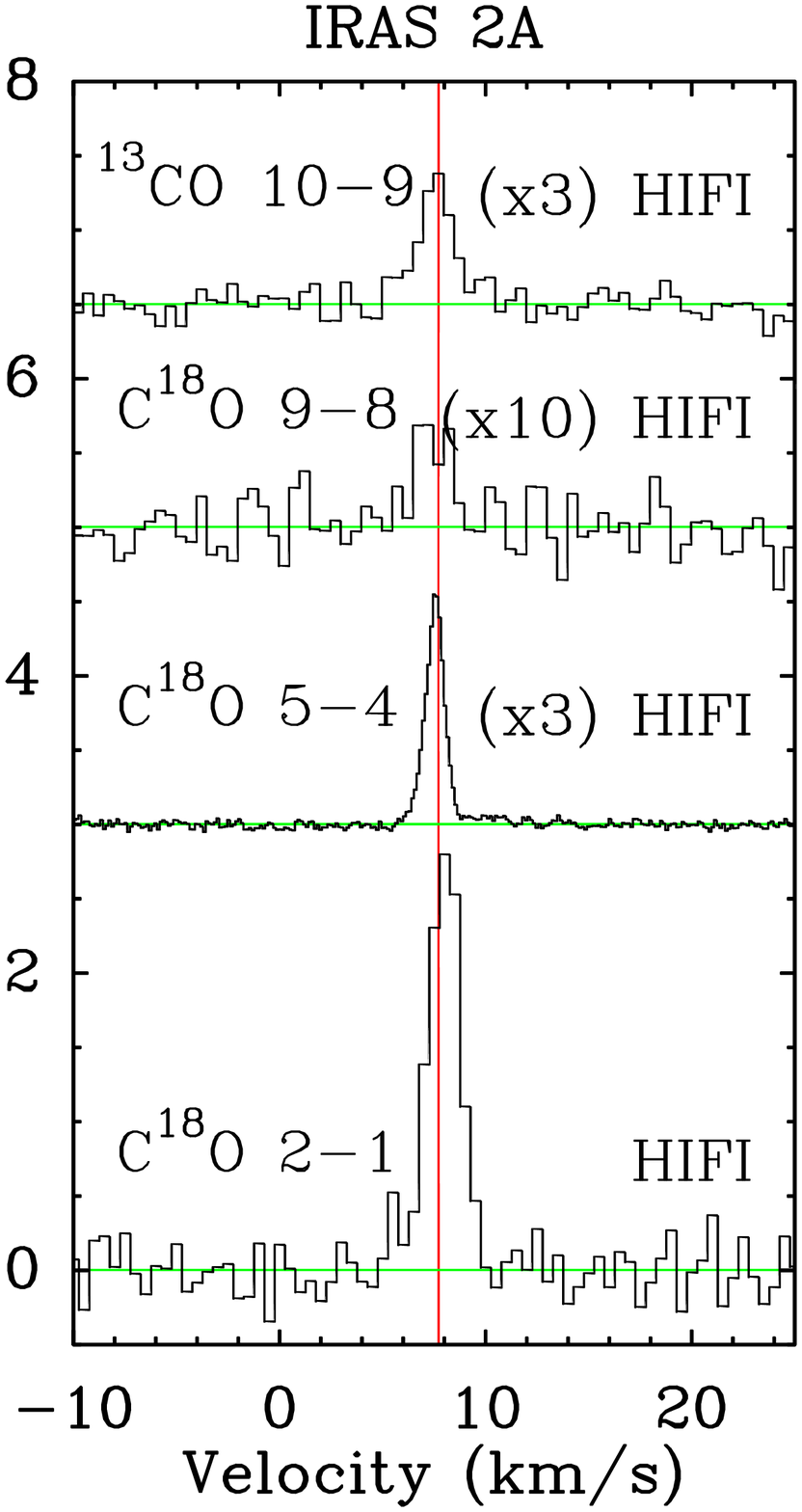}
\includegraphics[scale=0.36]{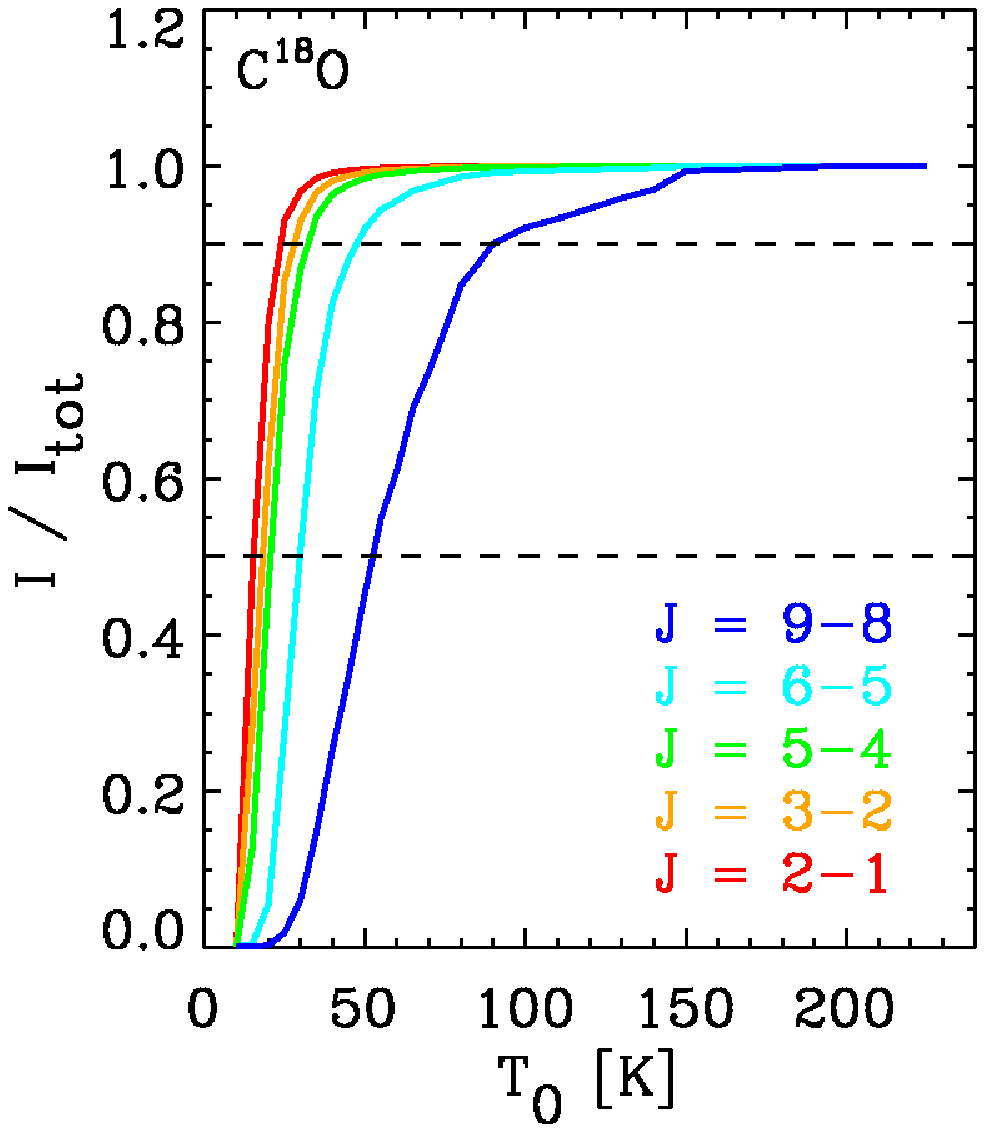}
\includegraphics[scale=0.27]{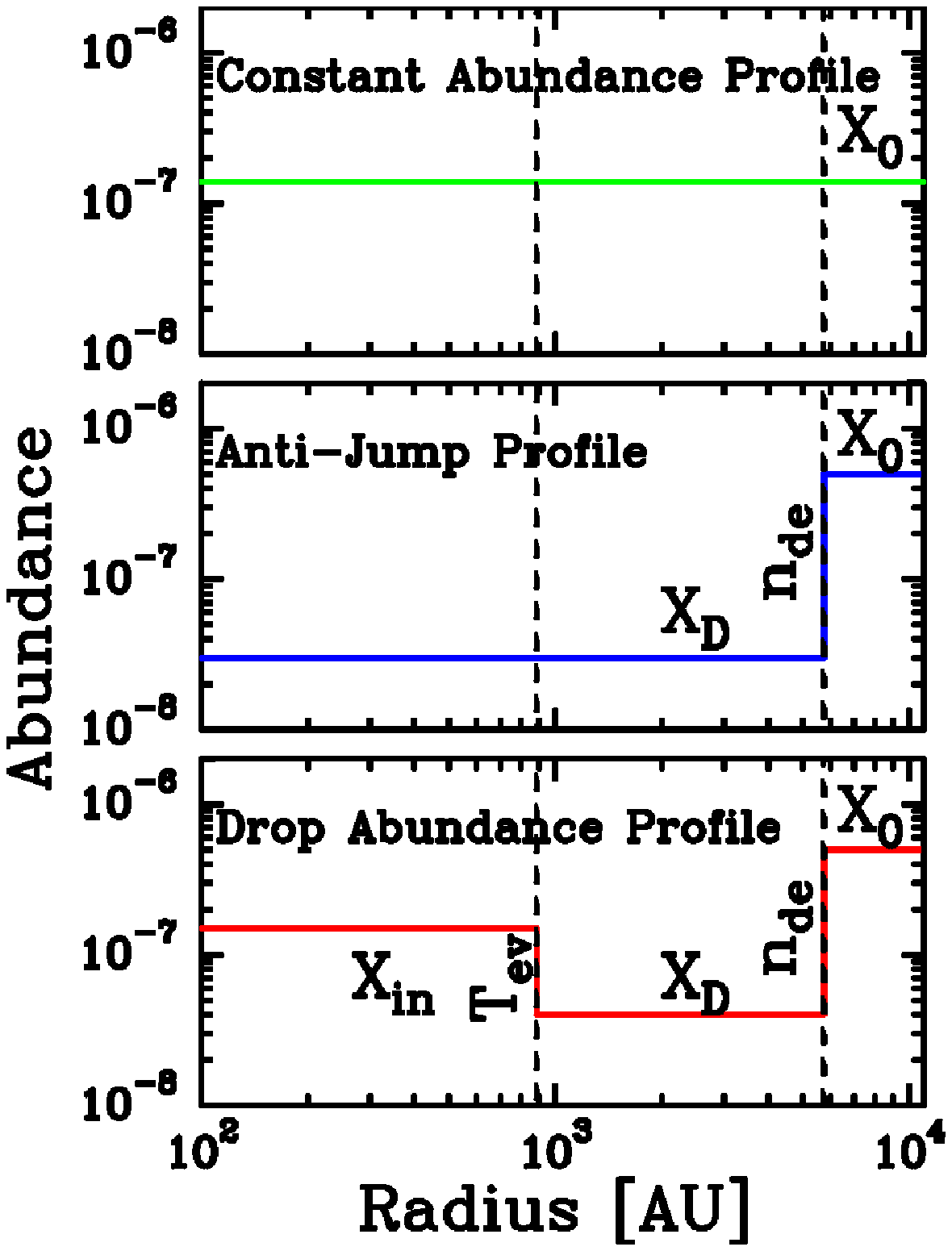}
\end{center}
\caption{\scriptsize ({\it Left}:) IRAS~4A $^{12}$CO and IRAS~2A $^{13}$CO and C$^{18}$O, ground-based and {\it Herschel} spectra 
are shown. ({\it Middle}:) Each curve represents 
the fraction of the line intensity for a given transition which has its origin in gas below $T_{0}$. 
({\it Right}:) Constant, anti-jump and drop abundance model for IRAS~2A.}
\label{figs}
\end{figure}

Fig. \ref{figs} ({\it left}) presents the CO spectra from {\it Herschel} and from the ground. These data show;  
({\it \bf i}) $^{12}$CO line widths increase towards higher-$J$. 
({\it \bf ii}) CO and its isotopologues trace different material. The $^{12}$CO~10--9 lines are dominated by broad 
(FWHM 25--30~km~s$^{-1}$) emission indicating a large-scale shock process ($>$1000~AU). 
%Radiative transfer models are used to constrain the temperature of this shocked gas to 100--200 K. 
Several CO and $^{13}$CO line profiles also reveal a medium-broad component 
(5--10~km~s$^{-1}$), seen prominently in H$_2$O lines (\cite{Kristensen10}), indicative of small-scale shocks ($<$1000~AU). Narrow C$^{18}$O lines 
(2--3~km ~s$^{-1}$) trace the quiescent envelope material.
%Column densities for both components are presented, providing a reference 
%for determining abundances of other molecules in the same gas. 
% namely; $^{12}$CO traces shocked gas on larger scales , $^{13}$CO traces 
%small-scale shocks ($<$1000~AU) and C$^{18}$O traces the quiescent envelope material, 
%({\it \bf iii}) Temperature for the outflowing gas was determined in the CO line 
%wings from 6--5/10--9 ratio, implying a range of $T\sim$70--200~K.
({\it \bf iii}) Higher-$J$ isotopologue lines (e.g., \mbox{C$^{18}$O~9--8} and 10--9; $E_{\rm up}$=237 and 289~K) trace high temperatures. However, analysis shows that for a power-law envelope $\sim$50\% of the observed emission still comes from the colder $<$40~K  parts of the envelope (Fig. \ref{figs} {\it middle}). Three different models 
using the \verb1RATRAN1 radiative transfer code are applied to the optically thin 
C$^{18}$O lines to quantify the CO abundance. For the ``constant'' profile, the observed line intensities 
cannot all be reproduced simultaneously. For the ``anti-jump'' model, the outer abundance, 
$X_{0}$, was kept high for densities lower than $n_{\rm desorption}$ (\cite{Jorgensen05}). 
It fits well the lower-$J$ lines, but the higher-$J$ 
lines are underproduced. In the ``drop'' model, the inner abundance, $X_{\rm in}$, increases 
above $T_{\rm ev}$. However, $X_{\rm in}$ is a factor of $\sim$3-5 below $X_{0}$ (Fig. \ref{figs} {\it right}). 
Best fit to the IRAS~2A data is found with this model
%Full description of the modelling and the comparison with the observed lines are given in 
(see \cite{yildiz2010} for details).

%%-----------------------------
%%      your bibliography
%%-----------------------------

\end{document}